\documentclass[12pt]{article}
\usepackage{graphicx}
\begin{document}
{\Large Efficient Hopfield pattern recognition on a scale-free neural network}

\bigskip
Dietrich Stauffer$^{1,2,4}$, Amnon Aharony$^{1}$, Luciano da Fontoura Costa $^3$
and Joan Adler $^4$

\bigskip
\noindent
$^1$ School of Physics and Astronomy, Raymond and Beverly Sackler Faculty of
Exact Sciences, Tel Aviv University, Ramat Aviv, Tel Aviv 69978, Israel

\medskip
\noindent
$^2$ Institute for Theoretical Physics, Cologne University, 
D-50923 K\"oln,  Euroland

\medskip
\noindent
$^3$ Cybernetic Vision Research Group, IFSC-USP, Caixa Postal 369, 
13560-970 S\~ao Carlos, SP, Brazil

\medskip
\noindent
$^4$ Department of Physics, Technion-IIT, Haifa 32000, Israel
 
\medskip
\noindent
e-mail: aharony@post.tau.ac.il, stauffer@thp.uni-koeln.de,

luciano@if.sc.usp.br, phr76ja@techunix.technion.ac.il

\bigskip
{\small  
Neural networks are supposed to recognise blurred images (or patterns)
of $N$ pixels (bits) each. Application of the network to an initial blurred
version of one of $P$ pre-assigned patterns should converge to the correct
pattern. In the ``standard" Hopfield model, 
the $N$ ``neurons'' are connected to
each other via $N^2$ bonds which contain the information on the stored patterns.
Thus computer time and memory in general grow with $N^2$. The Hebb rule assigns
synaptic coupling strengths proportional to the overlap of the stored patterns
at the two coupled neurons. 
Here we simulate the Hopfield model on the Barab\'asi-Albert scale-free network,
in which each newly added neuron is connected to only $m$ other
neurons, and at the end the number of neurons with $q$ neighbours decays
as $1/q^3$. 
Although the quality of retrieval decreases for small $m$, we find good
associative memory for $1 \ll m \ll N$. Hence, these networks gain a
factor $N/m \gg 1$ in the computer memory and time. 
}

PACS: 05.40-a, 05.50+q, 87.18.Sn

\bigskip
Traditional neural network models have nodes $i$ (``neurons'') coupled to all
other nodes $k$ with some coupling constant $J_{ik}$ (''synaptic strength''),
similar to Sherrington-Kirkpatrick infinite-range 
spin glasses \cite{sk}. Here we consider one of the simplest neural
network models, due to Hopfield \cite{hopfield}. This model was mostly 
applied to infinite range and was only rarely
put onto a square lattice with short-range interactions \cite{forrest,kurten}.
Real neural networks seem to have neither infinite nor only nearest-neighbour 
connections. The spatial structures of neural networks were investigated 
\cite{luciano} and compared with small-world and scale-free networks
\cite{ba,watts,benjacob,morita,karbowski}. Now we present computer simulations
of the Hopfield model \cite{hopfield} with Hebb couplings between neighbours 
restricted to a Barab\'asi-Albert (BA) scale-free network \cite{ba}.

In the Hopfield model, each of $N$ neurons or sites can be firing ($S_i=+1$) 
or not firing ($S_i=-1$). Neurons are coupled through $J_{ik}$, 
and are sequentially updated according to 
$$S_i \rightarrow {\rm sign} (\sum_k J_{ik} S_k) \quad . \eqno(1)$$
(We mostly ignore the diagonal terms $i=k$ in our sums.)
This rule corresponds to a low-temperature 
Monte Carlo simulation of a spin glass. The model has stored $P$ different
patterns $\xi_i^\mu \; (\mu=1,2\dots,P),$ which we take as random strings of
$\pm 1$. The couplings are given by the Hebb rule: 
$$ J_{ik} = \sum_\mu \xi_i^\mu \xi_k^\mu \quad .\eqno (2) $$ 
The first of these patterns is presented to the network in a 
corrupted form $S_i$, with ten percent of the $S_i$ reversed in comparison to 
the correct $\xi_i^1$. The question is whether the iteration through Eq. (1)
transforms the erroneous $S_i$ into the correct $\xi_i^1$. The quality of this 
pattern recognition is given by the overlap
$$ \Psi = \sum_i S_i \xi_i^1/N \quad ,\eqno (3)$$
which is related to the Hamming distance and equals 1 for complete recognition
and $\sim \pm 1/\sqrt N$ for only accidental agreement at random sites; it is 
$\sim 0.8$ at the beginning of the pattern recognition process, due to the ten 
percent reversal. 

Now we restrict the synaptic connections $J_{ik}$ to neurons which are 
neighbours in the BA network, but we still use Eqs. (1-3). In
these networks, we start from a small core of $m$ sites, all connected with 
each other. Then $N \gg m$ nodes are added, one after the other. Each new site 
$i$ selects exactly $m$ sites from the already existing network sites as its
neighbours $k$, with a probability proportional to the number of neighbours
which the existing site $k$ has already: The rich get richer. When the network
has added $N$ sites with a total of $N+m$ sites, its growth is stopped and the 
neural process of Eqs. (1-3) starts. Synaptic connections $J_{ik}$ exist only 
between sites $i$ and $k$ which are neighbours.
 
Since no longer every neuron is connected to all other neurons, the 
memory-saving trick of Penna and Oliveira \cite{penna} to avoid storing the
$J_{ik}$ no longer applies.
400 Megabyte were needed for $N=10,000$ nodes and $P=20,000$ patterns.
To save computer time, the $J_{ik}$ should be determined {\it after} and not
before the construction of the BA network.

When only one pattern is stored, it is recognised completely after two
iterations. With $P > 1$, however, no complete recognition takes place,
the overlap $\Psi$ is usually at the final fixed point (reached after about
five iterations) lower than at the beginning, as shown in Fig. 1a . However, the
model can still recognise the first pattern as the one presented to it, since
the overlap $\Psi \sim 0.19$ for $P=N=10^4$ is still appreciably larger 
than the overlap $|\Psi| < 500$ with the other $(P-1)$ patterns. 

Rather similar results are obtained if we work on a nearest-neighbour hypercubic
lattice with $N=L^d$ sites, similar to the studies made in 
\cite{forrest,kurten} in two dimensions.
Fig.1b shows that only for small numbers $P$ of patterns an increased $d$ means
an increased final overlap. For $d=7, \; 10$ and 15 the overlaps with up to 20 
patterns did not differ appreciably from $d=5$. No significant size effects were
seen for $4 \le L \le 20$ in five and $4 \le L \le 13$ in seven dimensions. 

Fig. 1 is based on one sample only for each point; using instead 100 samples at 
$m=3$ and $N=10,000$, we see in Fig. 2 that the overlap varies roughly as
$\Psi(P) - 0.19 \propto P^{-0.6}$, except for very small $m$. 
A similar power law $P^{-0.6}$ is also
found for hypercubic lattices (not shown.) 
It would be interesting to understand this power law from some analytical 
analysis.

A much better recovery of the corrupted pattern is obtained if we take a larger 
inner core of the BA network, that means if $m$ is no longer
small. (The first $m$ network sites are all mutually connected, as in the 
traditional Hopfield model.) Using 100 patterns, Fig. 3a shows the overlap 
$\Psi$ for $N+m = 10^4$
total sites as a function of $m$. Already at $m=200, \; N=9800$ the final
overlap is 88 percent; at $m = 2000, \; N = 8000$ we have complete recovery.
For $ 1 \ll m \ll N$ the number of connections (counting each 
bond twice) is $mN$ in our case and $N^2$ in the fully connected case;
thus we saved connections by a factor $m/N$. If we include the diagonal terms 
$J(i,i) = P$ in Eq. (1), we prevent the overlap from
becoming worse than the initial overlap 0.8 for small $m$ and still get overlap
near unity for large $m$, Fig. 3b.

Of course, with a large $m$ the network is no longer scale free, as shown in
Fig. 4: The simple power law $\propto 1/q^3$ for the number of sites with $q$ 
neighbours each \cite{ba} persists for $10^2 < q  <10^3$, but
a Gaussian peak is added for large $q$. However, the additional bump
concerns only a relatively small number of sites, and is probably
negligible for any practical purposes.
 
For infinite range, $m=N$,
the usual Hopfield model \cite{volk} gives an overlap $\Psi$ close to $1$ for 
$P/N < 0.14$ and a relatively small overlap $\Psi \sim 0.2$ for $P/N > 0.14$, 
with a sharp jump at $P/N = 0.14$. Our simulations, in contrast, show a gradual
deterioration as soon as more than one pattern is stored, but the value of 
$\Psi$ is still of order 0.2 and distinctly larger than for the other $(P-1)$ 
patterns. Using a medium-sized fully connected core, like $m \sim 10^3 $ at 
$P \sim 10^2$, surrounded by a larger BA network with $N \sim 10^4$ sites, gives
a good compromise between good recovery and not too many connections. 
 
So far it is not clear if the efficient
recovery simply results from the relatively large average coordination
number $m$, 
or by some additional ingredients in the
problem. It would also be interesting if Nature takes advantage of a similar 
efficiency. If it does, do natural neural networks share some geometrical 
features with the large-$m$ (but finite) scale-free networks \cite{ba}? 

We thank D. Horn for a discussion and  the German-Israeli Foundation for
supporting our collaboration.

\newpage
\begin{figure}[hbt]
\begin{center}
\includegraphics[angle=-90,scale=0.45]{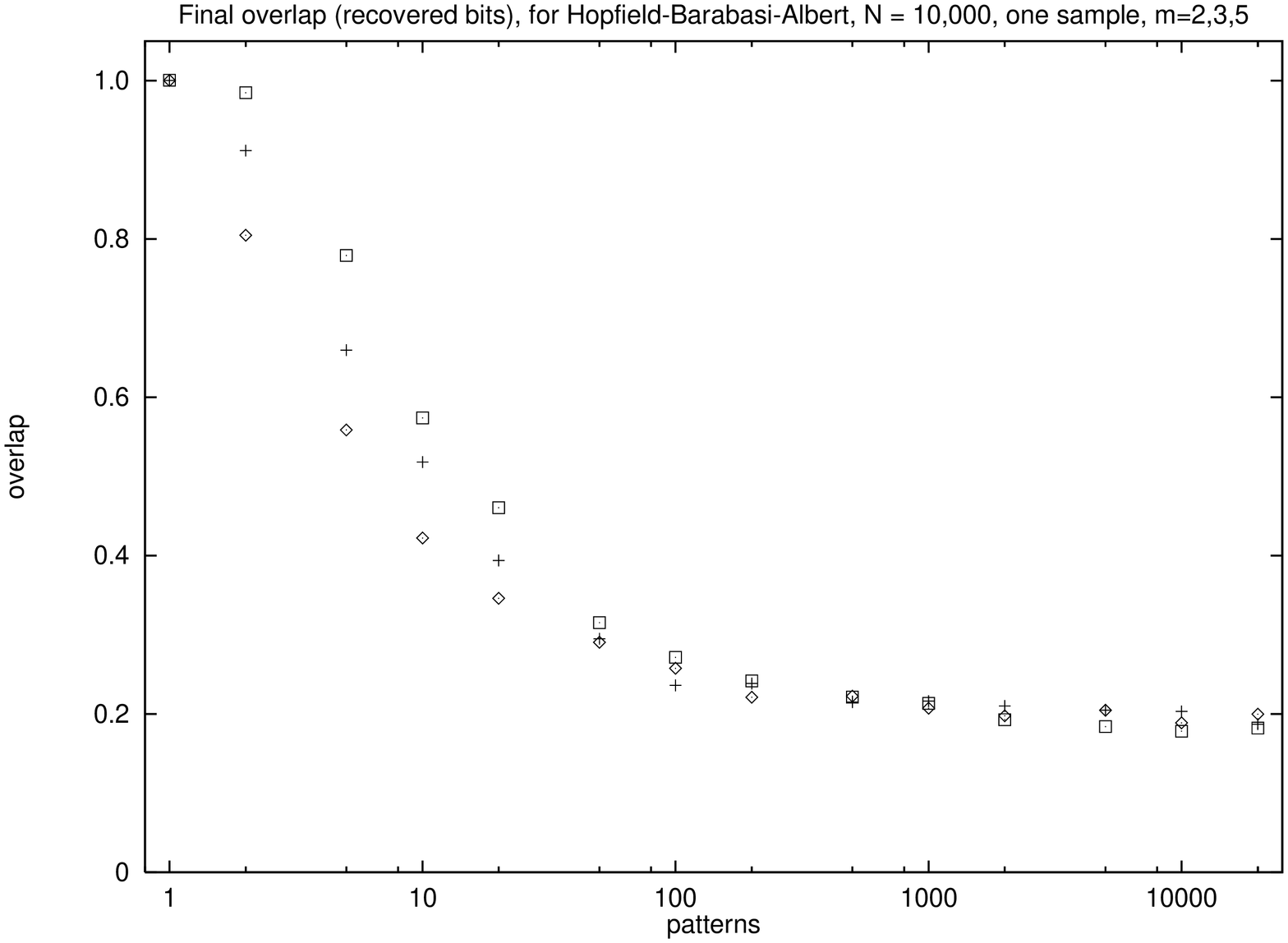}
\includegraphics[angle=-90,scale=0.45]{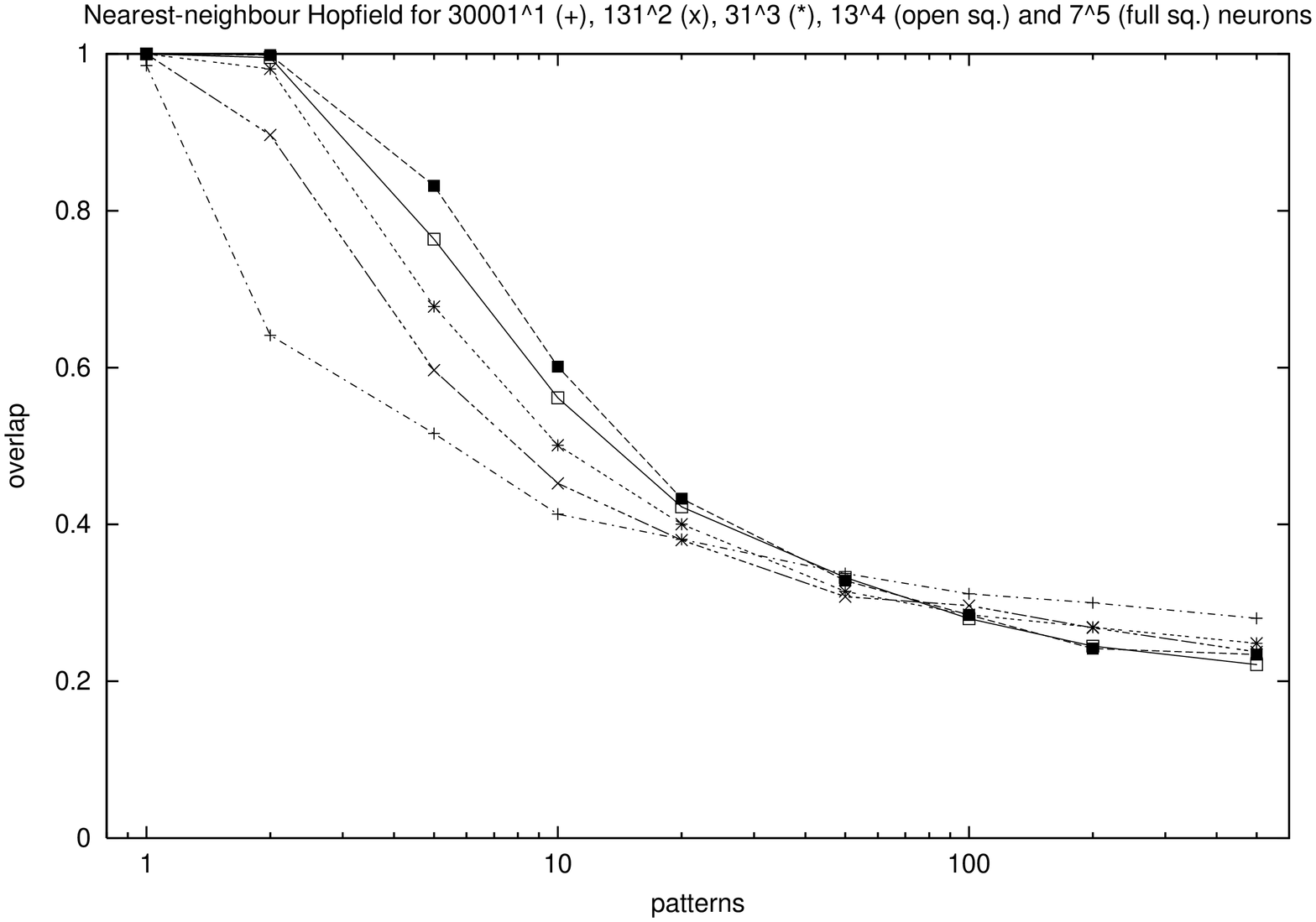}
\end{center}
\caption{ Final overlap $\Psi$ as a function of the number $P$ of patterns, for
$N \sim 10^4$ neurons. Each point is based on one sample only. Part a: BA 
network with $m=2,\; 3$ and 5 from bottom to top. Part b: Nearest-neighbour 
hypercubic lattice in one to five dimensions as shown in headline.
}
\end{figure}
 
\begin{figure}[hbt]
\begin{center}
\includegraphics[angle=-90,scale=0.50]{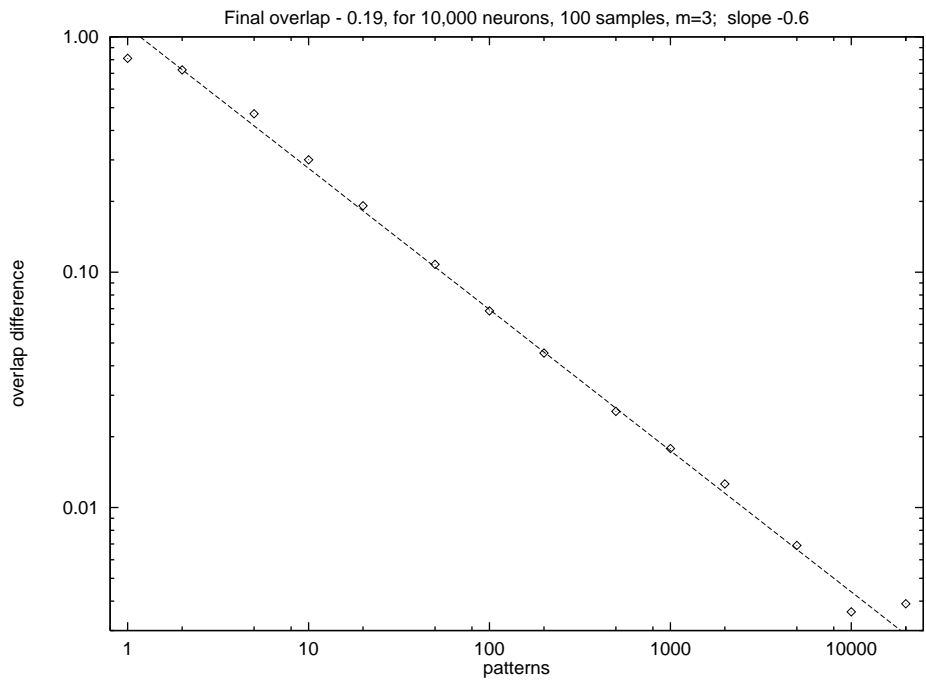}
\end{center}
\caption{ 
Approximate power law variation of final overlap difference ($\Psi-0.19$) with 
the number of patterns $P$, averaged over 100 samples, with $N = 10^4$ at $m=3$.
The straight line has a slope $-0.6$.
}
\end{figure}
 
\begin{figure}[hbt]
\begin{center}
\includegraphics[angle=-90,scale=0.41]{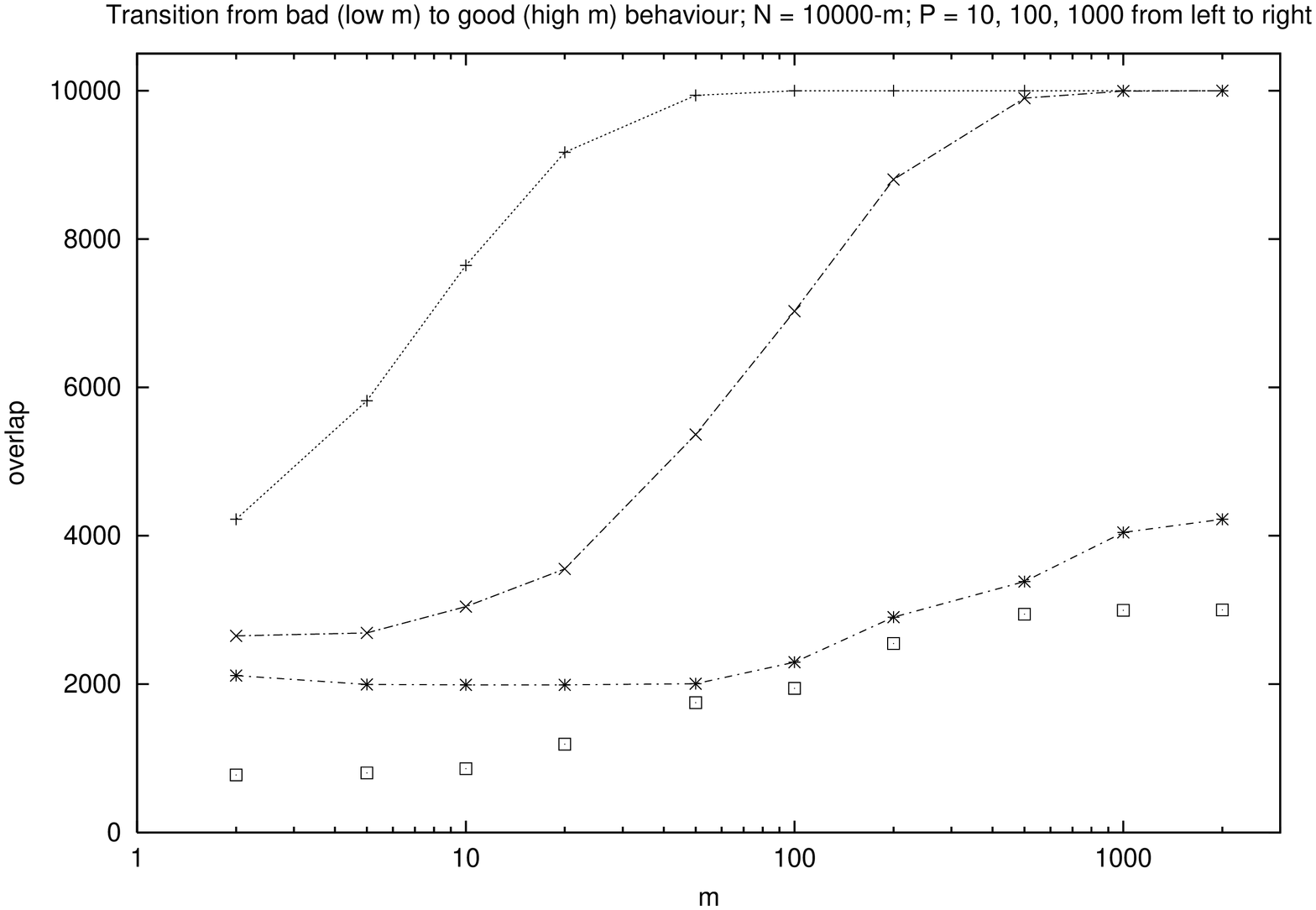}
\includegraphics[angle=-90,scale=0.41]{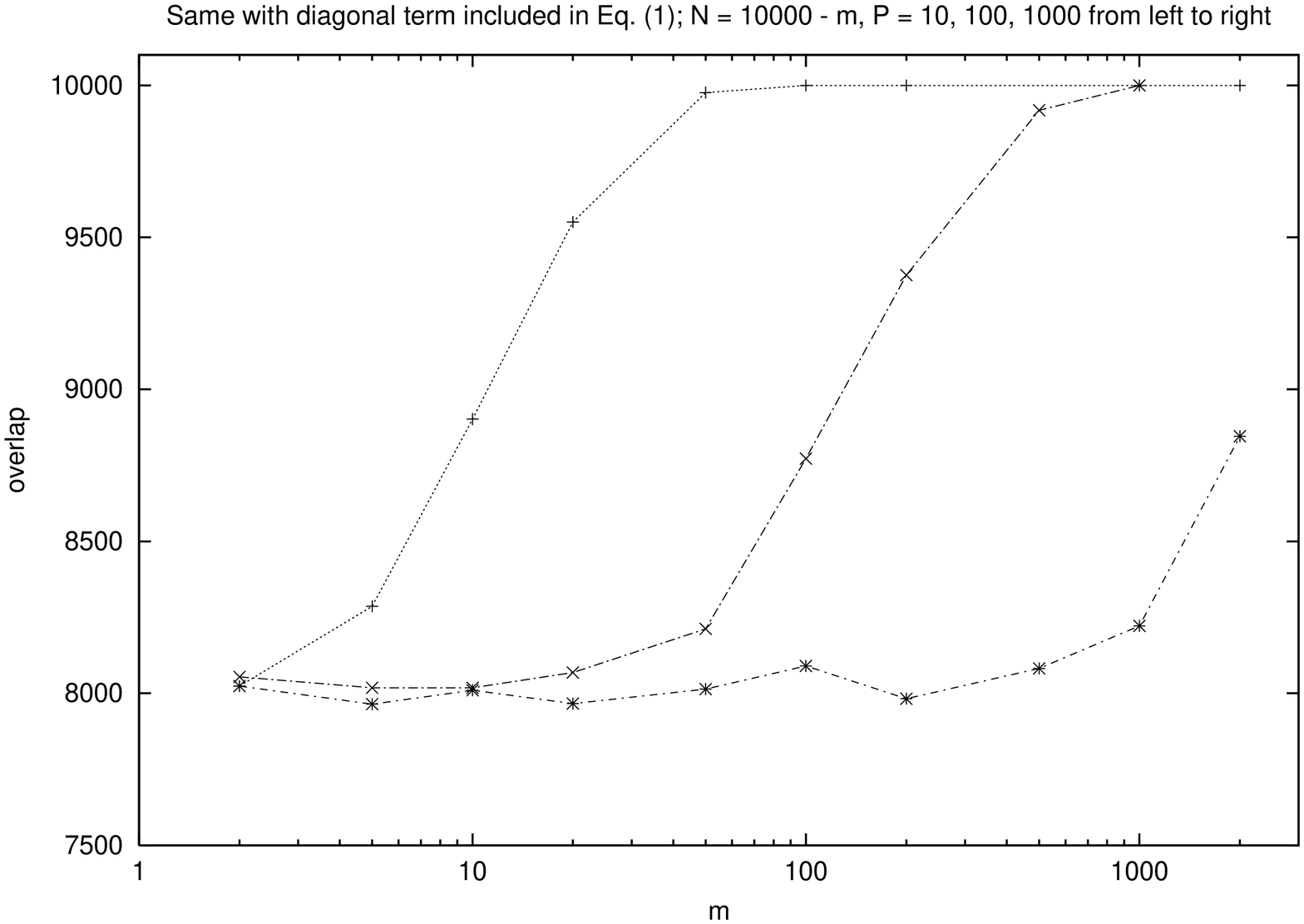}
\end{center}
\caption{ 
Variation of final overlap (not normalized) 
with the size $m$ of the fully connected core, 
surrounded by $N = 10^4 - m$ BA sites having $m$ neighbours each, at $P = 10$,
100 and 1000 (from left to right). Already for $P \ll m \ll N$ the corrupted 
pattern is restored well. The lowest  
data points refer to $P=100, \; N = 3000-m.$ Part a ignores the diagonal term
in the sum (1), while part b includes it. 
}
\end{figure}
 
\begin{figure}[hbt]
\begin{center}
\includegraphics[angle=-90,scale=0.50]{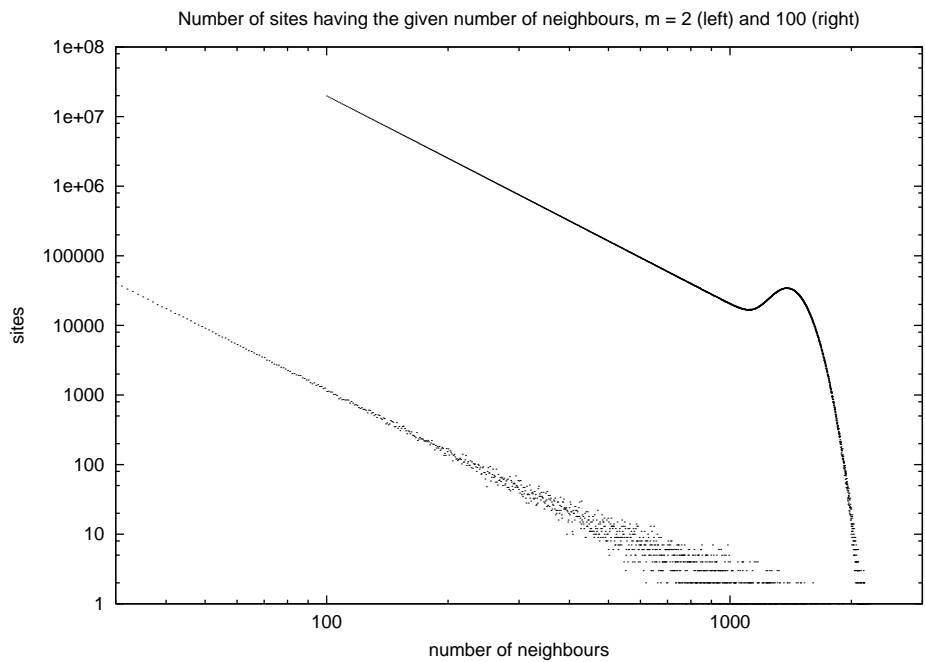}
\end{center}
\caption{ 
Number of sites having $q$ neighbours for $N = 10^4$, summed over 100,000 
simulations, at $m=100$ (right data). We no longer get the simple power law 
const/$q^3$, shown here for comparison at $m=2$ (left data).
}
\end{figure}
 
\end{document}